\documentclass[fleqn,twoside]{article}
\usepackage{espcrc2}

\usepackage{graphicx}
\usepackage[figuresright]{rotating}

\newcommand{\AmS}{{\protect\the\textfont2
  A\kern-.1667em\lower.5ex\hbox{M}\kern-.125emS}}
\newcommand{\tr}{^{\rm tr}}

\newcommand{\U}{{\rm U}}
\newcommand{\SU}{{\rm SU}}
\title{BCS Diquark Condensation in the $3+1$d Lattice NJL Model}

\author{David N. Walters \address{Department of Physics,
        University of Wales Swansea, \\
	Singleton Park, Swansea SA2 8PP, United Kingdom.}
        }

\begin{document}

\begin{abstract}
We present preliminary evidence of BCS diquark condensation in the $3+1$ dimensional
Nambu--Jona-Lasinio (NJL) model at non-zero chemical potential ($\mu$)
on the lattice.
Large $N$ results are used to match the model's parameters to  low energy, zero
density phenomenology.
A diquark source $j$ is added in a partially quenched approximation to
enable the measurement of lattice diquark observables. 
In particular measurements are made of the diquark condensate and
susceptibilities as functions of $j$ which support the existence of a
BCS phase at high $\mu$.
\vspace{1pc}
\end{abstract}

\maketitle

\section{Introduction}

Despite recent advances in the study of QCD at small but non-zero
density on the lattice \cite{lowmu} the sign problem still persists in the
study of dense, low temperature, {\em exotic} matter phases. One approach is to study
four-fermi theories which, unlike $\SU(2)_C$ QCD
(in which the $qq$ baryons are bosonic), can display
 Fermi-surface effects such as BCS superfluidity.
Although these models exhibit no
confinement, one variant, the $\SU(2)_L\times\SU(2)_R$ NJL model,
possesses the same 
global symmetries as 2 flavour QCD, which should allow us to investigate
the phase structure of this theory. 

\section{Lattice Model and Parameter Choice}

The model studied here, which uses staggered lattice fermions,
 is the 3+1 dimensional version of that studied
in \cite{NJL3} and references therein. In particular the model has the action
\begin{equation}
S={\bf \Psi}\tr {\cal A} {\bf \Psi} 
	+\frac{2}{g^2}\sum_{\tilde{x}}\left(\sigma^2 +
	\vec{\pi}.\vec{\pi}\right),
\end{equation}
where the bispinor ${\bf \Psi}$ is
written in terms of independent isospinor fermionic fields via
${\bf \Psi}\tr =(\overline{\chi},\chi\tr)$, and the auxiliary bosonic
fields $\sigma$ and $\vec{\pi}$ are introduced in the standard way.
 Written in the Gor'kov basis the fermion matrix is
\begin{equation}
{\cal A}=\frac{1}{2}\left(
\begin{array}{cc}
\overline{\jmath}\tau_2 & M \\ -M\tr & j\tau_2
\end{array}
\right),
\end{equation}
where the matrix $M$ is identical to the $3+1$d form of that given in \cite{NJL3}.
The diquark sources $j$ and $\overline{\jmath}$ differ from those
in \cite{NJL3} by a factor of 2 which allows us to identify them with
Majorana masses.

Being a theory of point-like interactions the NJL model has no
continuum limit for $d\geq4$ leaving the physics sensitive to 
simulation parameters. 
In order to ensure we simulate in a physical regime
these parameters are fitted to physical observables calculated in the 
large $N$ limit at $\mu=0$ \cite{SPK}. In particular we evaluate
dimensionless ratios between the pion decay rate $f_\pi$, the
constituent quark mass $m^*$ and the pion mass $m_\pi$, and fit to
the phenomenological values of 93MeV, 350MeV and 138MeV respectively.
Using these fits we determine the bare quark mass $am_0=0.002$ and the
inverse coupling $a^2/g^2=0.565$. We also extract the large $N$
lattice spacing $a^{-1}=1076$MeV, which allows us to present results
in physical units. 

\section{Preliminary Results}
The above model was simulated with a Hybrid Monte Carlo algorithm on
$L_s^3\times L_t$ lattices with $L_s=L_t=12$, $16$ and $20$. A
partially quenched approximation was used in which the diquark sources
are set to zero during the update of the auxiliary fields but are non-zero
during the measurement of diquark observables.

In order to study chiral symmetry breaking we measure the chiral
condensate $\left<\overline{\chi}\chi\right>$ and baryon number
density $n_B$, which are defined by
\begin{equation}
\left<\overline{\chi}\chi\right>= 
	\frac{1}{V}\frac{\partial\ln{\cal Z}}{\partial m_{_0}}
\end{equation}
and
\begin{equation}
n_B = \frac{1}{2V}\frac{\partial \ln {\cal Z}}{\partial \mu}.
\end{equation}

\begin{figure}[h]
\centering
\includegraphics[width=7.5cm]{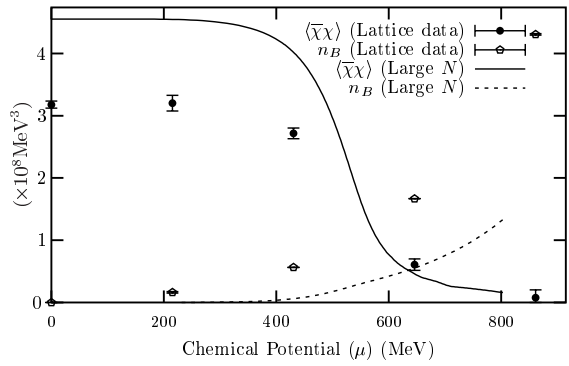}
\caption{Chiral condensate $\left<\overline{\chi}\chi\right>$ and
 baryon number density $n_B$ as functions of $\mu$ with lattice results
compared to the large $N$ limit.}
\label{EofS}
\end{figure}
Plots of $\left<\overline{\chi}\chi\right>$ and $n_B$ as functions of
chemical potential are presented in Figure \ref{EofS}. Both the large
$N$ solutions and the lattice data have been extrapolated to the limit
$V^{-1}\rightarrow 0$. 
Chiral symmetry is broken in the low $\mu$ phase, as
signalled by an non-zero chiral condensate, and is approximately 
restored as $\mu$ is increased through what appears to be a
crossover. The order of the chiral 
phase transition in the NJL model is strongly dependent on the
parameters used \cite{SPK}.

In the diquark sector, the operators
\begin{equation}
qq_\pm(x)=\chi^{tr}\frac{\tau_2}{4}\chi (x) 
\pm\overline{\chi}\frac{\tau_2}{4}\overline{\chi}^{tr}(x) 
\end{equation}
allow one to define the diquark condensate as
\begin{equation}
\left<qq_+\right>=\frac{1}{V}\frac{\partial\ln{\cal Z}}{\partial j_+},
\end{equation}
where $j_\pm=j\pm\overline{\jmath}$.

\begin{figure}[ht]
\centering
\includegraphics[width=7.5cm]{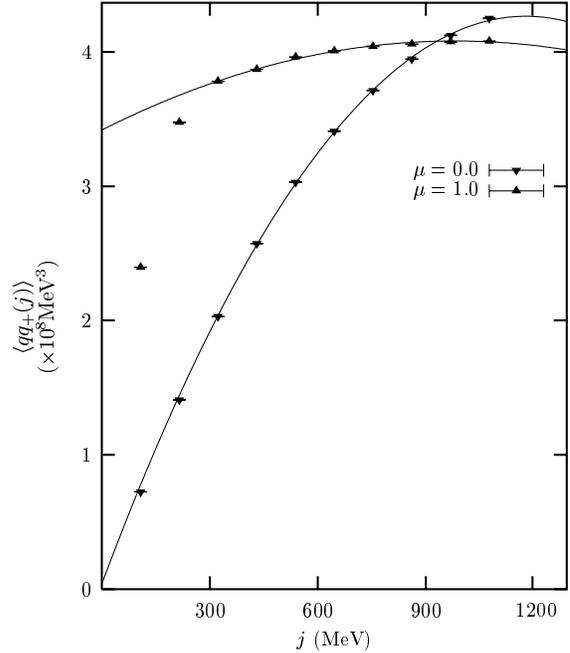}
\caption{Diquark condensate vs. $j$ for high and low $\mu$. }
\label{qq}
\end{figure}
Figure \ref{qq} shows the condensate plotted as a function of $j$
extrapolated to $L_t^{-1}\rightarrow 0$ (we simulate with
$j=\overline{\jmath}$ throughout). For $\mu=0$ a
quadratic fit through the data is consistent with the condensate vanishing as
$j\rightarrow 0$. For high $\mu$ the fit is consistent
with a non-zero diquark condensate which is similar in magnitude to the chiral
condensate in the vacuum. We believe that finite volume effects cause
the points with $j\le 0.2$ to fall below the curve. In particular,
in the high $\mu$ phase as $j\rightarrow 0$ we 
expect an exact Goldstone mode which leads the correlation length to
diverge. For this reason these points have been ignored during the fits.

Finally, if we define susceptibilities
\begin{equation}
\chi_{_\pm}=\sum_x\left<qq{_\pm}(0)qq_{_\pm}(x)\right>
,\end{equation}
we can clearly distinguish between the two phases by studying the
ratio
\begin{equation}
R={\displaystyle
\lim_{j_+\rightarrow0}}-\frac{\chi_{_+}}{\chi_{_-}}
.
\label{R}
\end{equation}
If a $\U(1)$ baryon number symmetry is manifest,
the two susceptibilities should be identical up to a sign factor and
the ratio should equal 1.
If the symmetry is broken, the Ward identity
\begin{equation}
\left.\chi_-\right|_{j_-=0}=\frac{\left<qq_+\right>}{j_+}
\end{equation}
predicts that $\chi_-$ should diverge and the ratio should vanish.

\begin{figure}[h]
\centering
\includegraphics[width=7.5cm]{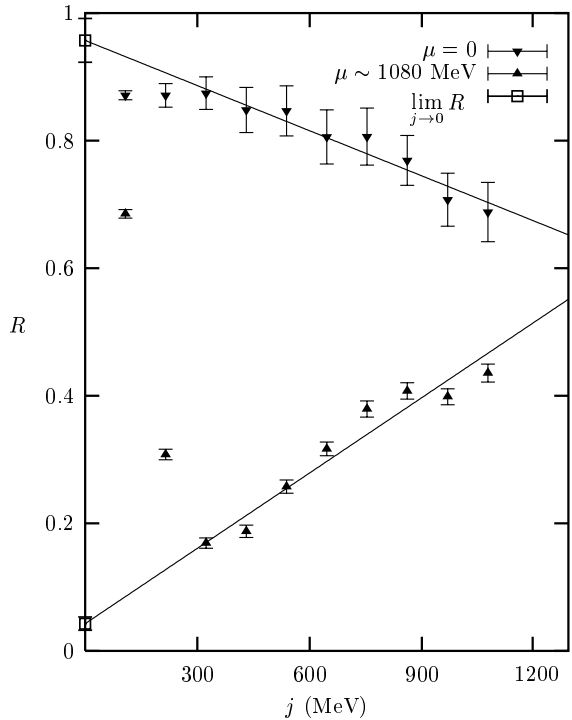}
\caption{Susceptibility ratio $R$ vs. $j$ for high and low $\mu$.}
\label{Rat0}
\end{figure}
Figure \ref{Rat0} shows the susceptibility ratio as a function of $j$, 
 again extrapolated to  $L_t^{-1}\rightarrow 0$. A
linear fit through the $\mu=0$ data is consistent with a
manifest $\U(1)_B$ symmetry. For $\mu a=1.0$ we see markedly
different behaviour, with a fit (again discarding $j\le 0.2$) suggesting
an intercept close to zero and therefore a broken baryon number
symmetry.

\section{Summary and Outlook} 
We have provided preliminary evidence of a BCS superfluid phase
in the 3+1d NJL model at high
$\mu$, with an order parameter approximately equal in magnitude to the
chiral order parameter at $\mu=0$.
  The behaviour of the susceptibility ratio
defined in (\ref{R}) further supports a broken
$\U(1)_B$ symmetry in the high $\mu$ phase analogous to the broken
$\U(1)_{EM}$ in BCS superconductors.   
This study provides the first evidence of a BCS phase in lattice field
theory, which we interpret as evidence of a
colour-superconducting phase at high $\mu$ in QCD.

Clearly, more data are required for intermediate values of $\mu$ to  
understand the nature of the diquark transition. To have 
more control over the finite volume effects shown in Figures \ref{qq}
and \ref{Rat0} it may also be necessary to study lattices with $L_s\neq L_t$.
Finally, in the future we wish to study the fermion dispersion
relation. This will allow us to present a comparison between the
chiral mass gap $\Sigma$ and the BCS superfluid gap $\Delta$ which
can, in principle, be experimentally determined.

\section*{Acknowledgements}
This project is being carried out in collaboration with Simon Hands.

\end{document}